\documentclass[12pt,english,a4paper]{article}
\usepackage{amssymb,amsmath}
\usepackage[dvips]{graphicx}
\usepackage{epsf}
\usepackage{babel}
\usepackage{float}

\begin {document}
\large { \vspace * {6.0cm}
\begin {center}

{ \Large
 A. A. Vasenko, N. D. Galanina, K. E. Gusev, V. S. Demidov,\\
E. V. Demidova, I. V. Kirpichnikov, A. Yu. Sokolov,\\
 A. S. Starostin, and N. A. Khaldeeva

\vspace {2.0cm}
{ \bf
 Formation of $^{24}$Mg$^*$ in the Splitting of
$^{28}$Si Nuclei by 1-GeV Protons }}

\vspace {2.0cm} Institute of Theoretical and Experimental Physics,\\
ul. Bol'shaya Cheremushkinskaya 25, Moscow, 117218 Russia

\end {center}

{ \normalsize

The $^{28}$Si(p,p$^{\prime}\gamma_{0}$X)$^{24}$Mg reaction has been
studied at the ITEP accelerator by the hadron-gamma coincidence
method for a proton energy of 1~GeV. Two reaction products are
detected: a 1368.6-keV $\gamma$-ray photon accompanying the
transition of the $^{24}$Mg$^*$ nucleus from the first excited state
to the ground state and a proton p$^{\prime}$ whose momentum is
measured in a magnetic spectrometer. The measured distribution in
the energy lost by the proton in interaction is attributed to five
processes: the direct knockout of a nuclear $\alpha$ cluster, the
knockout of four nucleons with a total charge number of 2, the
formation of the $_{\Delta}$Si isobaric nucleus, the formation of
the $\Delta$ isobar in the interaction of the incident proton with a
nuclear nucleon, and the production of a $\pi$ meson, which is at
rest in the nuclear reference frame. The last process likely
corresponds to the reaction of the formation of a deeply bound pion
state in the $^{28}$P nucleus. Such states were previously observed
only on heavy nuclei. The cross sections for the listed processes
have been estimated. }

\clearpage

\large

\begin {center}
{\bf 1. INTRODUCTION.}
\end {center}

Stodolsky~[1] proposed to use the coherent properties of an atomic
nucleus for the suppression or enhancement of various nuclear
reaction mechanisms. The experimental verification of theoretical
predictions~[1] is difficult, because it is necessary to separate
nuclear states with excitation energies $\sim$MeV when incident
particle energies are$\sim$GeV.

As shown in~[1-6], the fixation of the residual nucleus state allows
the effective separation of certain reaction channels. Some
reactions can be treated as semicoherent, because certain
transitions are associated with the collective excitations of a
certain group of particles.

In the MAG magnetic germanium spectrometer~[7], used in this
experiment, prompt $\gamma$-rays from the transitions of excited
nuclei to a state with a lower excitation energy or to the ground
state are detected. This makes it possible to determine the state of
the residual nucleus including its quantum numbers.

In view of this circumstance, investigation of the formation of
$^{24}$Mg in the collision of protons with $^{28}$Si nuclei by the
hadron-gamma coincidence method at the MAG spectrometer is of great
interest, because the process can be treated in terms of the
$\alpha$ cluster model, the $^{28}$Si target nucleus is the
$^{12}$C--$\alpha$--$^{12}$C nuclear molecule~[8]. In this model,
the quasi-free particle can be a target for the incident proton and
provide the direct knockout of particles from the nucleus in
peripheral interactions.

Central interactions of protons with the nucleus can lead to
collective excitations whose energies are higher than the discrete
levels and giant resonances~[9]. The average excitation energy of
the residual nuclei formed after completion of the nuclear cascade
induced by 1-GeV protons in the $^{28}$Si is $\sim$0.06~GeV and the
excitation energy distribution expands to 0.4~GeV or higher.

The reaction
\begin{equation}
^{28}\mbox{Si}(p,p^{\prime}\gamma_{0}X)^{24}\mbox{Mg} \mbox{ , }
\label{eq:1}
\end{equation}
that is more general than the quasielastic knockout of particles was
investigated at the MAG spectrometer. Here, p$^{\prime}$ is the
secondary leading proton whose momentum is measured by the magnetic
spectrometer, $\gamma_{0}$ is the photon accompanying the transition
of the $^{24}$Mg nucleus from the first excited state to the ground
state (its energy is used to identify the final nuclear state), and
$X$ is any combination of secondary hadrons (nucleons and pions)
with a total atomic number of 4 and a total charge number of 2. The
direct knockout reactions correspond to $X=\alpha$. Competitive
cascade interactions of the proton with quasi-free nucleons in the
nucleus that lead to the emission of four nucleons ($X$=$2p2n$) are
simultaneously detected. In this work, reaction (1) is investigated
in a wide energy-transfer range up to 0.8~$T_0$ , where $T_0$ is the
proton beam energy. This range includes not only the excitations of
discrete levels but also $\pi$-meson and isobar nuclear excitations.

\begin {center}
{\bf 2. EXPERIMENTAL PROCEDURE AND DATA PROCESSING }
\end {center}

The experiment was carried out with the universal beam of the ITEP
proton synchrotron for $T_0$=1.02~GeV. Reaction (1) was studied by
the hadron-gamma coincidence method using the MAG magnetic germanium
spectrometer~[7] The silicon target ($^{28}$Si--92.23\%,
$^{29}$Si--4.67\%, $^{30}$Si--3.1\%) has a thickness of
$x$=6.18~g/cm$^2$. Gamma-ray photons from the target were detected
by the germanium spectrometer based on a Ge(Li) crystal and the
leading secondary charged particles (predominantly hadrons) were
detected by the magnetic spectrometer based on a wide- aperture
magnet. The track part of the spectrometer consists of multiwire
proportional chambers placed in front of and behind the magnet. The
momenta $p$ of the charged particles were measured in the magnetic
spectrometer with the accuracy $\frac{dp}{p}=(0.32p+0.57)\%$, where
$p$  is measured in GeV. Angles were measured with an accuracy of
0.003~rad. The germanium spectrometer was adjusted for detecting
prompt photons emitted from the target perpendicularly to the beam
direction and for measuring their energy from 50 to 3000~keV. We
selected 892 596 events with a $\gamma$-ray photon and one charged
hadron emitted from the target at angle $\theta<15^{\circ}$  to the
beam direction. In the photon-energy distribution of the events, a
number of lines corresponding to certain $\gamma$-transitions in the
reaction product nuclei are observed against the background. Twenty
eight $\gamma$-transitions are identified for 19 product nuclei~[10,
11]. To analyze reaction (1) of the formation of $^{24}$Mg*, we
considered range {\bf A} of the $\gamma$-ray energies 1350-1380~keV
including the line $E_{\gamma 0}$=1368.6~keV of the transition of
the $^{24}$Mg nucleus from the first excited state to the ground
state.

Two factors distort the angular distributions of secondary protons
detected in the magnetic spectrometer. The first factor is
associated with random events induced by beam protons that do not
interact in the tar- get. Small angles $\theta$ are characteristic
of these events. The second factor is the dependence of the
detection efficiency of the magnetic spectrometer on $\theta$. This
factor is associated with the limited sizes of track detectors and
their rectangular geometry. This factor is noticeable for
$\theta>6.5^{\circ}$  and, if necessary, is taken into account by
correcting coefficients.

\begin {center}
{\bf 3. DATA ANALYSIS FOR ${\bf 3^{\circ}<\theta<6.5^{\circ}}$.}
\end {center}

In this region, the random-particle background can be disregarded
and the efficiency of the magnetic spectrometer can be taken as
$\approx$ 100\%. Figure 1 shows the fragment of the energy spectrum
of photons for these events near the $E_{\gamma 0}$, line whose
tabulated value is indicated by the arrow. The histogram part
corresponding to range {\bf A} are doubly shaded. The number of
events in this range is $N_2$=1965. The right and left shaded
intervals {\bf B}, are used to determine the background under peak.
The spectrum fragment is approximated by the sum of a normal
distribution with the standard deviation equal to the resolution of
the $\gamma$-spectrometer~[7, 11] and a linear function describing
the background. The least squares fit are shown by the smooth
curves. The number of events of reaction (1) in interval {\bf A} is
$N_3$=745$\pm$73 whereas the background from the continuous is
$N_{b}$=1217.

Figure 2 shows the distribution of events of reaction (1)in the
energy $\omega=T_0+M_p-\sqrt{M_p^2+P^2}$ ($M_p$ -- is the proton
mass and $P$ -- is the scattered-proton momentum measured in the
spectrometer) that is transferred by the proton interacting with the
target nucleus. This distribution is obtained by subtracting the
background from the $\omega$ distridution of $\omega$$N_2$ events in
interval {\bf A}. The background distribution in $\omega$ is
determined in intervals {\bf B} and is normalized to $N_b$ events.

The peak near $\omega\approx0$ corresponds to the interaction of
beam protons with quasi-free $\alpha$ clusters. Negative values
appear due to the errors of the measurement of the momentum $P$ in
the magnetic spectrometer. Events in the range
$\omega$=0.22--0.6~GeV, where the production of mesons is
energetically possible, are attributed to the formation of the
isobar and isobaric nuclear excitation. The shape of the
distribution of events with the maximum at
$\omega_0$~$\approx$0.145~GeV is characteristic of the production of
a pion that is at rest in the laboratory system or in the system
associated with the nucleus.

In order to describe the experimental data, the following processes
are simulated:\\
1) proton scattering on the nuclear $\alpha$ cluster with the
emission of the cluster from the nucleus,\\
2) the formation of the $_{\Delta}$Si isobaric nucleus,\\
3) the formation of the $\Delta_{1232}$ isobar in the interaction of
the beam proton with a nuclear nucleon, and\\
4) the production of a pion that is at rest in the system
associated with the nucleus.\\

The $\omega$ distributions of events for these processes are
calculated by the Monte Carlo method using the GEANT-3.21 program.
The conditions of the detection of particles in the magnetic
spectrometer and the selection of events with respect to the proton
scattering angle are reproduced in the simulation program. The
momentum of one leading particle, which is emitted from the nucleus
at angle $3^{\circ}<\theta<6.5^{\circ}$ and has positive charge, is
determined. The possibility of multiple rescattering of the proton
with the highest energy in the nucleus is taken into account for
processes 1 and 3. The ratio between the probabilities of various
multiple scatterings for $^{28}$Si is calculated by the Monte Carlo
method using the mean free path of the proton in the nucleus and
known cross section for the nucleon-nucleon inter- action. The ratio
1~:~0.75~:~0.24 is obtained for the single, double, and triple
interactions, respectively. The Fermi momenta of the nucleon and
$\alpha$ cluster are taken into account when simulating the
interactions of the incident proton with the nuclear nucleon or
$\alpha$ claster. The mean Fermi momenta of the nucleon and cluster,
which are determined by the separation energies for these particles,
are equal to 0.19 and 0.3~GeV/c, respectively. To ensure the energy
conservation in the reaction, the effective target mass at the
collision instant is calculated from the condition that the total
energy of the target is equal to its rest mass and the momentum is
taken as a random value from the Rayleigh distribution.

The conditions under which reaction (1) is energetically possible
are satisfied in simulation. Events for process 1 are selected such
that the first elastic interaction of the proton occurs with the
nuclear $\alpha$ cluster, which acquires an energy sufficient for
the emission from the $^{28}$Si nucleus, whereas the energy acquired
by nuclear nucleons in the subsequent interactions with the proton
is insufficient for their emission. Process 2 is simulated by the
interaction of the proton with the $^{28}$Si nucleus with the
excitation of the nucleus to an energy of 0.294~GeV (the average
energy of the isobar excitation of the nucleon). The excitation
energy in the particular interaction is taken as a random value from
the Breit-Wigner distribution with width $\Gamma$=0.12~GeV in
agreement with the conclusions made in [12] that the excitation
energies of the free isobar and isobaric nucleus are equal to each
other. Process 3 is simulated using the channels of the formation
and decay of the isobar and their branching ratios presented in
[13-15]. In the first interaction of the incident proton with the
nucleus, the event selection algorithm for process 3 ensures the
energy transfer sufficient for the emission of four nucleons from
the nucleus. In the second and third interactions of the proton with
a nuclear nucleon, if they occur, events in which the energy
transferred to the target proton is lower than its separation energy
are selected.

The c.m. scattering angle $\psi$ in two-particle processes 1-3 is
simulated as follows. This angle in processes 1 and 2 is taken such
that the distribution in the 4-momentum transfer squared is
exponential with an argument of 10 and 30 for the $\alpha$-particle
and silicon nucleus, respectively. In isobar formation reactions 3,
it is assumed that the c.m. angular distribution at its formation
stage corresponds to $cos^{4}~\psi$~[14]. The isobar decays are
simulated as isotropic.

The estimate for the contributions of four indicated processes to
events $N_{3}$ is obtained by using least squares fit of the
experimental $\omega$ distribution to the function consisting of the
sum of the model distributions for the above-indicated processes.
The designed parameters are the numerical contributions of these
processes. The fit is performed without the normalization to the
number of events in the experimental histogram. The optimum numbers
of events of processes 1- 4 calculated for the minimum
$\chi^{2}$=103, are equal to 461$\pm$41, 132$\pm$20, 45$\pm$12 and
27$\pm$8, respectively, for the number of degrees of freedom $n$=80.
The average $\omega$ value for a group of events responsible for
process 2 is equal to 0.145~$\pm$~0.003~GeV. The distribution is no
wider than the apparatus error so that the physical width of the
line is likely no larger than 0.004~GeV. Figure 2 shows the fitting
results. The statistical confidence of the existence of process 4
with the $\pi$-meson is estimated as 3.5 standard deviations.

\begin {center}
{\bf 4. ANALYSIS OF THE RESULTS FOR ${\bf
0^{\circ}<\theta<15^{\circ}}$.}
\end {center}

In this range, 7040 events accompanied by photons whose energies are
in interval {\bf A} were detected. The statistical treatment of
events in this range has three features. First, as shown in [7],
losses of events that are due to limited geometric sizes of the
coordinate detectors, their shape, and space arrangement cannot be
disregarded for $\theta<15^{\circ}$. Owing to these losses, the
efficiency of the magnetic spectrometer for detecting charged
particles is not equal to 1 and, moreover, depends on the angles and
momenta of the particles. The procedure for determining and taking
into account the efficiency $\varepsilon(\theta,p)$, was described
in [7]. According to this procedure, each detected event is assigned
with the weight $W=\varepsilon^{-1}(\theta,p)$ and the ''effective''
number of events is used in all numerical calculations reported
below. For the case under consideration, the sum of the weights of
events with $E_\gamma \in$~{\bf A} is equal to $W_1$=20~851.

Second, the feature of the procedure concerns the selection of
events of reaction (1): in addition to the linear background from
the continuous component of the photon spectrum, a noticeable
contribution from three additional sources exists in photon-energy
interval {\bf A}. These sources are $\gamma$-transitions in other
nuclei, random coincidences, and secondary interactions of the
products of reaction (1) with the target nuclei. The back- ground
from the continuous component and $\gamma$-transitions is
numerically determined by using of the least squares approximation
of the experimental distribution in $E_\gamma$ as in Section 3. The
contribution of secondary processes is calculated using the
reference data on the cross sections for the reactions of the
splitting of $^{28}$Si nuclei by protons and neutrons for various
energies. The number of random coincidences is determined using the
characteristic angular distribution measured in specially conducted
experiments. Table 1 presents the results calculated for the
background components.

{\bf Table 1. Background components\\
}
\begin{tabular}{|c|c|}
\hline Linear background,  $W_{b1}$ & 12748$\pm$600\\
\hline $\gamma$-transitions in other nuclei with photon
 & \\
energies in interval €,
$W_{b2}$ & 1241$\pm 250$\\
\hline Random-coincidence background & \\
from the beam protons, $W_{b3}$ & 825$\pm$125\\
\hline Background from the secondary processes, &\\
in the target, $W_{b4}$ & 697$\pm$154\\
\hline
\end{tabular}\\

After the subtraction of the background, $W_{(1)}$=5340$\pm$450
events of reaction (1) remains. Using this value, one can calculate
the cross section for reaction (1) for proton scattering angles
$\theta<15^{\circ}$ by the formula
\begin{equation}
\sigma_{(1)}=\frac{4\pi A W_{(1)}}{N_{A} x \Xi \nu \varepsilon_{\tau} k m g N_{0} \delta} \mbox{ ,}
\end{equation}
where $A$ is the atomic number of the $^{28}$Si nucleus, $N_{A}$ is
the Avogadro number, and $N_{0}$=2.36$\times$10$^{11}$ is the number
of protons incident on the target. The remaining coefficients are
presented in Table 2.

{\bf Table 2.  Coefficients appearing in Eq. (2)\\
 }
\begin{tabular}{|c|c|}
\hline Efficiency of the $\gamma$-detector &\\
at $E_{\gamma 0}$=1368.6~keV & $\Xi$=0.000777 \\
\hline Efficiency of the program reconstruction & \\
of one-track events & $\nu$=0.583\\
\hline Coefficient corresponding to miscounts & \\
due to the spread of the times of the arrival & $\varepsilon_{\tau}$=0.886\\
of signals from the germanium detector & \\
\hline Correction for the absorption of &\\
$\gamma$-rays in the target & $k$=0.614\\
\hline Correction for the dead time &\\
of the germanium detector & $m$=0.95\\
\hline Correction for the cutoff of the distribution  &\\
edges by the boundaries of interval {\bf €}  & $g$=0.979\\
\hline Correction for the percentage of &\\
$^{28}$Si in the target & $\delta$=0.922\\
\hline
\end{tabular}\\

The value $\sigma_{(1)}$=(10.1$\pm$0.9$\pm$1.5)~mb is obtained for
reaction (1) in the angular range $\theta<15^{\circ}$.

 The third feature of the processing procedure in this
range is that the description of the experimental distribution of
events $W_{(1)}$ in energy transfer $\omega$ (see histogram in Fig.
3) by the model curves corresponding to processes 1-4 is
supplemented by the following process:\\

5) the knockout of four nucleons with a total charge number of 2 in
the collision of the proton with a nuclear nucleon.\\

As processes 1 and 3, it is simulated with the inclusion of the
multiple rescattering of the proton with the highest energy in the
nucleus. The energy of separating four nucleons from the $^{28}$Si,
nucleus, which is assumed to be equal to 0.040~GeV, it is spent by
the proton in the first interaction event.

As a result of the approximation of reaction (1) by processes 1-5,
the minimum value $\chi^2$/$n$=98.9/83 is obtained. Table 3 presents
the numerical results of minimization. The first row presents the
contributions of processes in terms of the effective number $W$ of
events and the second row contains the cross sections for the
corresponding processes. The optimum calculation function is shown
by the solid line in Fig. 3 and its components are marked by the
respective numbers.

{\bf Table 3.  Partial cross sections for processes 1-5
\\ }
\begin{tabular}{|c|c|c|c|c|c|}
\hline & 1 & 2 & 3 & 4 & 5 \\
\hline $W$ &1259$\pm$214 & 763$\pm$162 & 1324$\pm$215 & 117$\pm$58 & 1652$\pm$247\\
\hline $\sigma$, mb &2.4$\pm$0.3 & 1.4$\pm$0.2 & 2.5$\pm$0.3 & 0.22$\pm$0.11 & 3.1$\pm$0.5\\
\hline
\end{tabular}\\

\begin {center}
{\bf 5. DISCUSSION}
 \end {center}

In this experiment, the cross section for reaction (1) and partial
cross sections for 1) proton scattering on the $\alpha$ cluster with
the emission of the cluster from the nucleus, 2) the formation of
the $_{\Delta}$Si isobaric nucleus, 3) the formation of the
$\Delta_{1232}$isobar, 4) the production of a pion that is at rest
in the system associated with the nucleus, and 5) the knockout of
four nucleons with a total charge number of 2 as a result of the
collision of the beam proton with a nuclear nucleon are measured
using the hadron-gamma coincidence method. The first and last
processes occur primarily at low energy transfers $\omega$ $<$
0.12~GeV, which are insufficient for the production of mesons. The
direct knockout of nuclear $\alpha$ clusters from the nucleus by
protons is well known both theoretically and experimentally [16,
17]. Although $\alpha$ particles are not detected in the MAG setup,
their emission is reliably identified, because the residual nucleus
$^{24}$Mg is determined (by $E_{\gamma 0}$) and energy transfer,
whose average value is equal to the $\alpha$ separation energy, is
measured. In contrast to the previous experiments, where the emitted
cluster is detected at certain angle $\theta_\alpha$, the method
used in this work allows measurement of the cross section for any
$\theta_\alpha$. The kinetic characteristics for process 5 of the
splitting of the $^{28}$Si nucleus with the emission of four
nucleons have not yet measured. Measurement of its probability at
the MAG setup became possible due to an experiment inclusive in
hadrons, where the distribution in the energy transfer $\omega$ was
analyzed.

The formations of the isobar and isobaric nuclei at energy $T_{0}
\approx$~1~GeV are the most probable processes in the energy
transfer range, where inelastic interactions with the production of
pions are energetically possible. This conclusion is based on the
experimental data for the interaction of protons with free nucleons.

The observation of events for $\omega \approx$ 0.145~GeV is likely
attributed to the formation of deeply bound $\pi^-$-mesonic atoms
$^{28}$P. Such states were predicted theoretically in [18] and were
observed experimentally in heavy nuclei Pb~[19], Xe~[20], and
Sn~[21]. Such phenomena have not yet observed in intermediate-mass
nuclei. Taking into account the threshold of the
p~+~$^{28}$Si~$\to$~p~+~$^{28}$P~+~$\pi^-$ reaction, which is equal
to about 0.154~GeV, the binding energy of the $\pi^-$ meson in the
phosphorus atom seems to be too high (0.009~$\pm$~0.003~GeV).
Another possible explanation is the formation of a deeply bound
state of the $\pi^0$ meson with the $^{24}$Mg, nucleus, which
appears after the knockout of the $\alpha$ particle from the
$^{28}$Si. nucleus by the incident proton. In this case, the binding
energy would be equal to 0.001--0.004~GeV.\\

We are grateful to G.A. Leksin, L.A. , A.B. Kaidalov, and Yu.A.
Simonov for discussion of the experimental results and their
interpretation. This work was supported by the Council of the
President of the Russian Federation for Support of Young Scientists
and Leading Scientific Schools (project no. NSh-1867.2003.2).\\

\vfill\eject

\begin{figure}[h]
\begin{center}
\mbox{ \epsfysize=17.0cm \epsffile{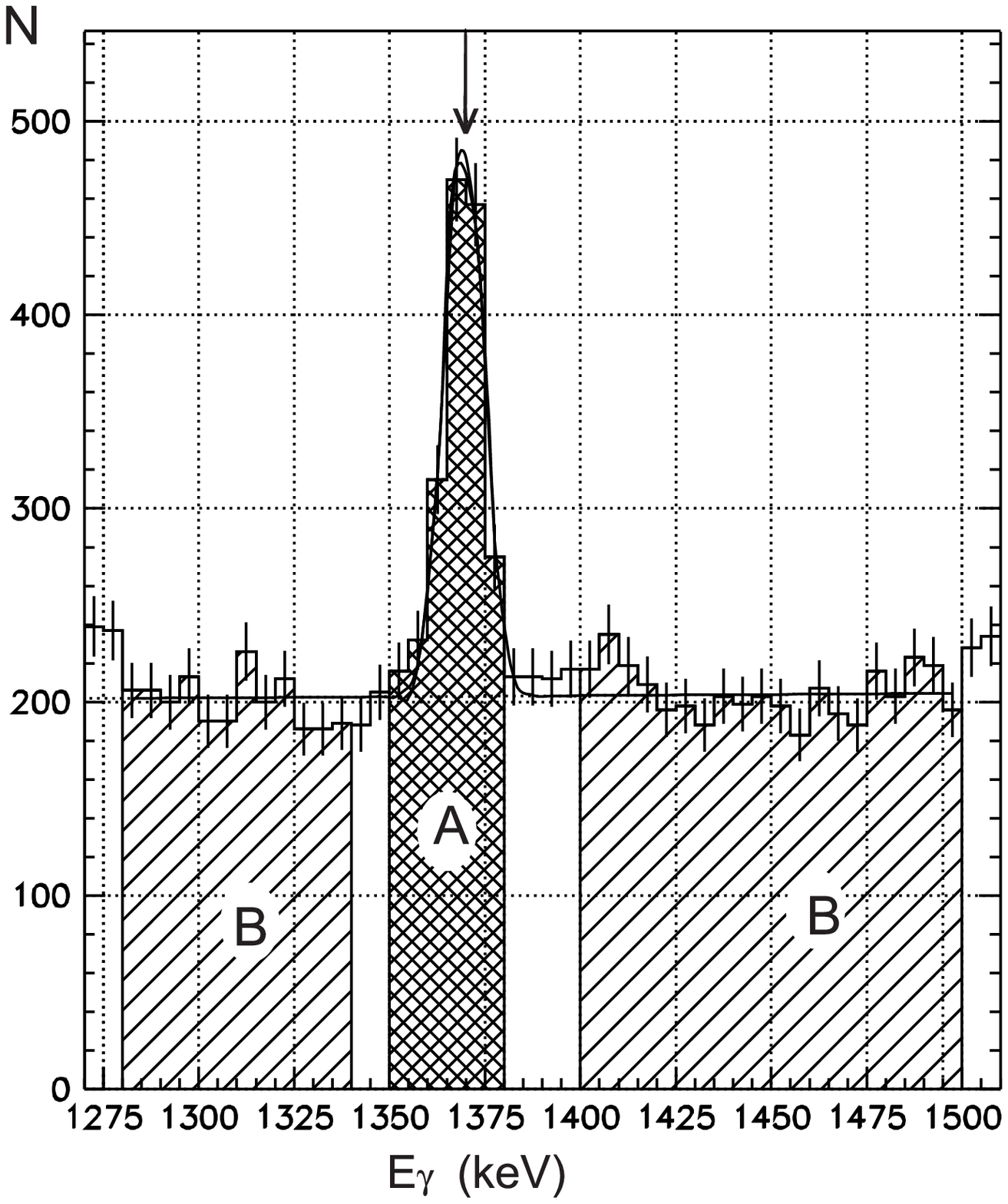} }
 \caption{Distribution
of events with proton emission angle $3^{\circ}<\theta<6.5^{\circ}$
in the $\gamma$--ray energy. Double shaded area {\bf A} is taken for
the selection of the $^{24}$Mg formation reaction. Shaded areas {\bf
B} are used to determine the background distribution shape. The
arrow indicates the tabulated energy of the $\gamma$ transition of
the $^{24}$Mg nucleus from the first excited state to the ground
state.}
\end{center}
\end{figure}

\begin{figure}[h]
\begin{center}
\mbox{ \epsfysize=17.0cm \epsffile{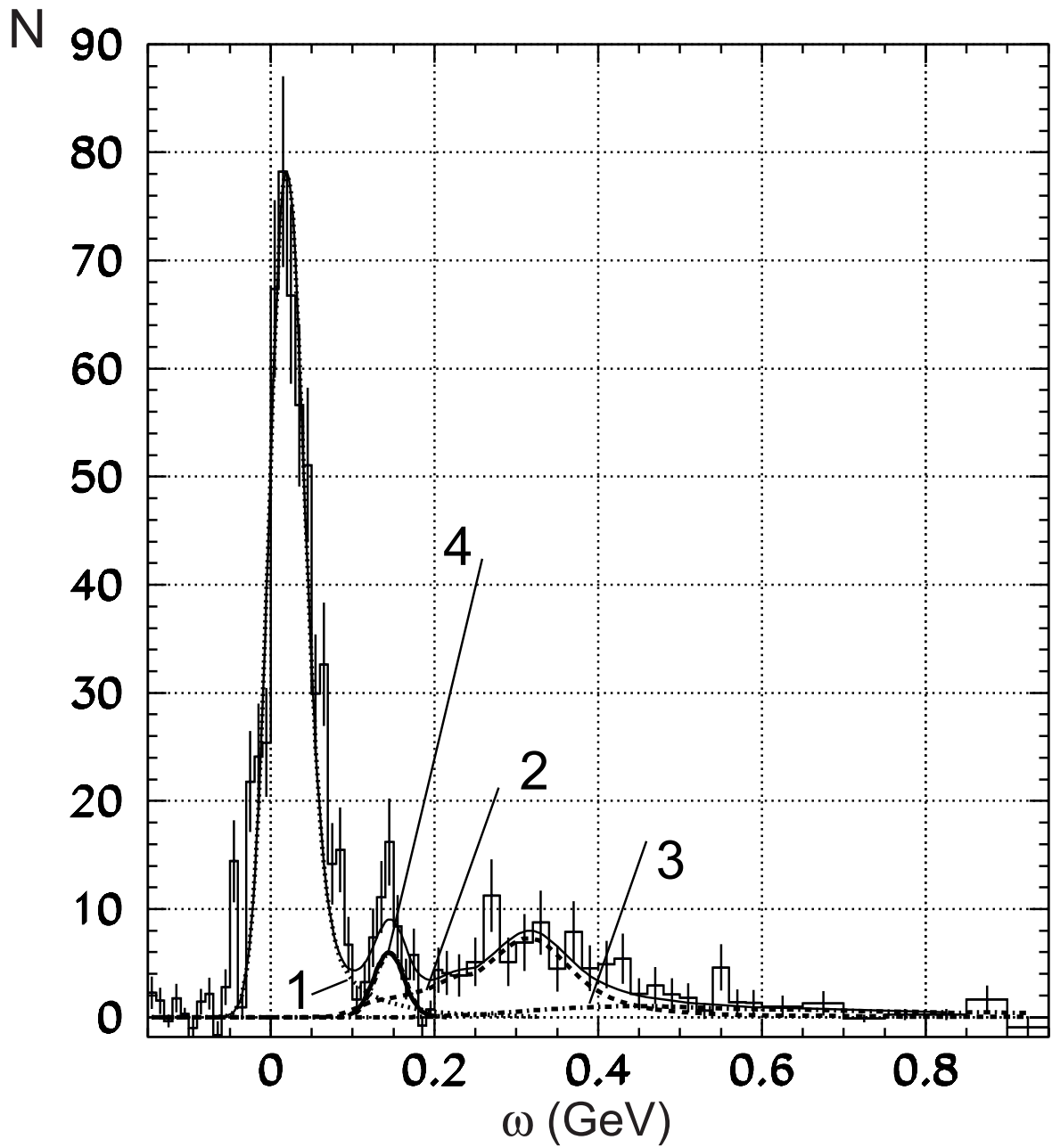} } \caption{
Distribution of events with angles $3^{\circ}<\theta<6.5^{\circ}$ in
the energy transfer $\omega$. The smooth lines are the fits by
processes 1-4 (see the main text).}
\end{center}
\end{figure}

\begin{figure}[h]
\begin{center}
\mbox{ \epsfysize=17.0cm \epsffile{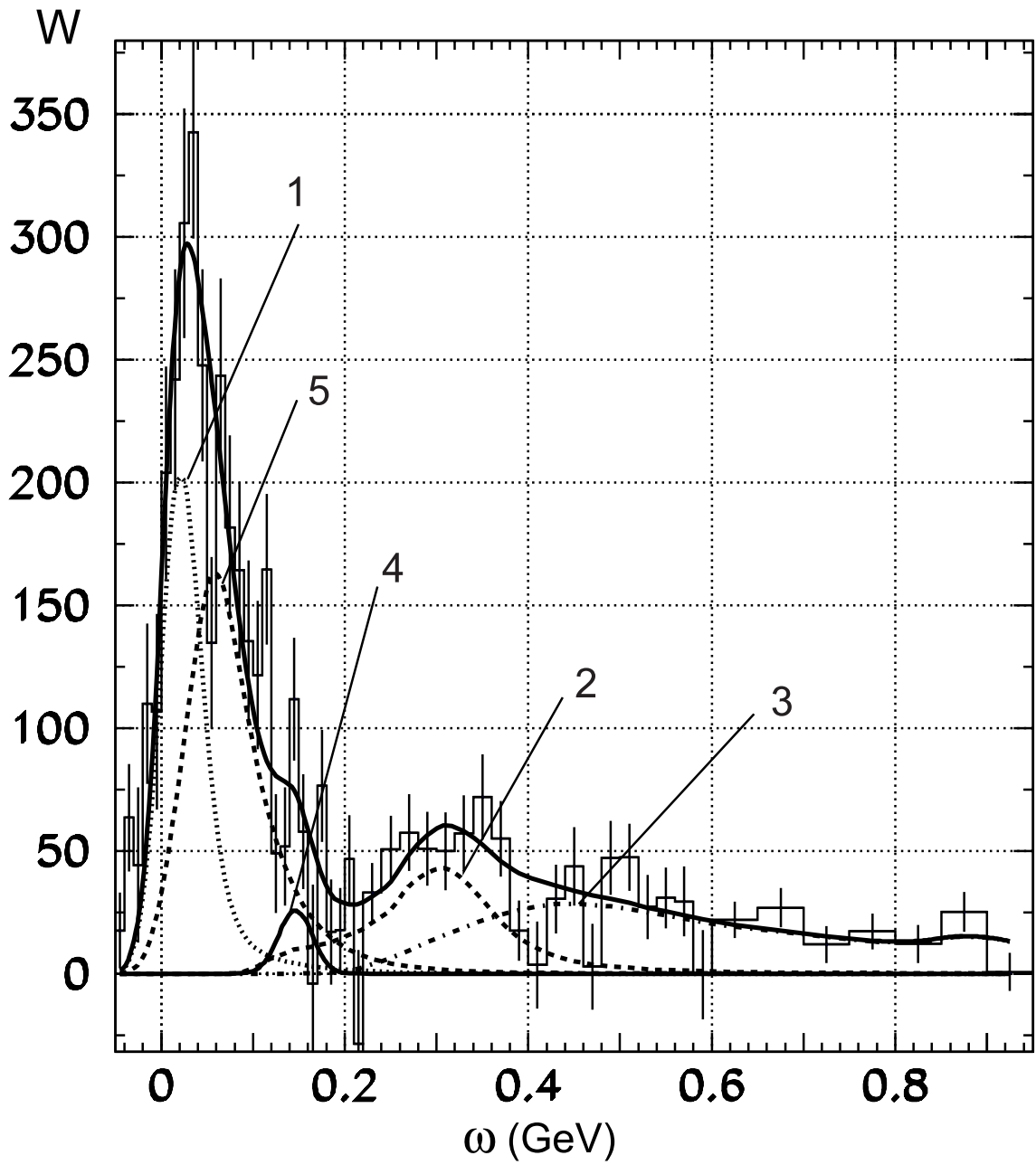} } \caption{
Distribution of weighted events with angles $\theta<15^{\circ}$ in
the energy transfer $\omega$. The smooth lines are the fits by
processes 1-5 (see the main text).}
\end{center}
\end{figure}

 }
\end{document}